# Speckle-based optical cryptosystem and its application for human face recognition via deep learning

Qi Zhao, [1,2,†] Huanhao Li, [1,2,†] Zhipeng Yu, [1,2,†] Chi Man Woo, [1,2] Tianting Zhong, [1,2] Shengfu Cheng, [1,2] Yuanjin Zheng, [3] Honglin Liu, [2,4] Jie Tian, [5,6*] and Puxiang Lai [1,2,7*]

[1]Department of Biomedical Engineering, Hong Kong Polytechnic University, Hong Kong SAR

[2]Shenzhen Research Institute, Hong Kong Polytechnic University, Shenzhen 518057, China

[3]School of Electrical and Electronic Engineering, Nanyang Technological University, Singapore

[4]Key Laboratory for Quantum Optics, Shanghai Institute of Optics and Fine Mechanics, Chinese Academy of Sciences, Shanghai 201800, China

[5]Beijing Advanced Innovation Center for Big Data-Based Precision Medicine, School of Medical Science and Engineering, Beihang University, Beijing 100191, China

[6]Key Laboratory of Molecular Imaging, Institute of Automation, Chinese Academy of Sciences, Beijing 100190, China

[7]Photonics Research Institute, Hong Kong Polytechnic University, Hong Kong SAR

[†]These authors contributed equally to this work.

[*]e-mail: tian@ieee.org; puxiang.lai@polyu.edu.hk

## Abstract

Face recognition has recently become ubiquitous in many scenes for authentication or security purposes. Meanwhile, there are increasing concerns about the privacy of face images, which are sensitive biometric data that should be carefully protected. Software-based cryptosystems are widely adopted nowadays to encrypt face images, but the security level is limited by insufficient digital secret key length or computing power. Hardware-based optical cryptosystems can generate enormously longer secret keys and enable encryption at light speed, but most reported optical methods, such as double random phase encryption, are less compatible with other systems due to system complexity. In this study, a plain yet high-efficient speckle-based optical cryptosystem is proposed and implemented. A scattering ground glass is exploited to generate physical secret keys of gigabit length and encrypt face images via seemingly random optical speckles at light speed. Face images can then be decrypted from the random speckles by a well-trained decryption neural network, such that face recognition can be realized with up to 98% accuracy. The proposed cryptosystem has wide applicability, and it may open a new avenue for high-security complex information encryption and decryption by utilizing optical speckles.

# Introduction

Human face is a personal identifier, and an adult can hardly change the appearance. In modern society, numerous face recognition scenes have been set up for authentication or security purposes due to the increasing concern for personal privacy and public safety[1]. The storage of human face data is hence highly confidential. If the face database is leaked, hackers may use this information to attack key sectors, including bank accounts[2]. Therefore, effective protection of face image data is essential for privacy and security[3].

Various cryptosystems, including software-based and hardware-based, have been put forward to protect private data. For software-based cryptosystems, well-known encryption algorithms have been developed, such as Rivest-Shamir-Adleman encryption (RSA)[4], Advanced Encryption Standards (AES)[5], Message Digest Algorithm (MD5)[6], *etc*. These algorithms are all based on mathematical theories whose digital secret key lengths range from tens to hundreds of bits. The selection of the secret key lengths involves a trade-off or balance between security level and processing speed. Such a limited key length seems to be sufficiently secure for conventional attacks by general computers but is vulnerable to attacks by the rapidly evolving quantum computers, whose computing power is $10^8$ times that of the general ones[7]. As a result, researchers keep exploiting novel cryptosystems to achieve higher security, and hardware-based solutions are therefore in demand.

Amongst current hardware-based solutions, optical cryptosystems are of extensive interest with the development of optical computing and computational imaging[8,9]. The optical methods may lead to breakthroughs in cryptosystem due to their superior performance, such as fast speed, high security, low cost, *etc*.[10]. Generally, optical cryptosystems use diffracted light to obtain the ciphertext from the plaintext (data or images to be encrypted), thus there is no computational cost and high-speed encryption (*i.e.*, speed of light) is guaranteed. Moreover, the large dimensionality of the optical diffraction mechanism guarantees a long length for digital secret keys, resulting in higher security[11]. In contrast, to achieve comparable secret key length in software-based cryptosystems, a high-performance computer is inevitable, and the cost is demanding. In view of these advantages, researchers have devised various optical cryptosystems, such as double random phase encryption (DRPE)[12,13] and speckle-based optical cryptosystems[14,15]. DRPE uses two phase masks at the input plane and the plaintexts are then encrypted on the Fourier plane. Although DRPE has been investigated for more than two decades, it is not yet widely adopted because it is difficult to be integrated with other systems.

Speckle-based approaches are therefore of interest, in which optical speckles are utilized as ciphertext to encrypt plaintext. Compared with DRPE, this method is much easier to implement with a plain optical setup. In a strong scattering regime, the plaintext (*e.g.*, images) is optically scrambled, resulting in speckles featured by randomly distributed bright and dark regions, which can be captured by regular digital cameras for further processing. The random feature of the speckles seems meaningless and usually annoying, but constitutes nearly infinite information channels[16] and hence tremendously long physical secret key length in a cryptosystem[12], which can be exploited to yield high-level security and information protection. Thus far, a few methods, such as based on transmission matrix[16,17], support vector regression[18], neural networks[14], *etc.*, have been developed to reconstruct images from the speckles. Among these approaches, neural networks can automatically learn the complex relationships between the plaintext and the ciphertext, resulting in image reconstruction of higher fidelity than other methods can yield[19-24]. Since the physical models in speckle-based optical cryptosystems are similar to those for imaging through scattering media, neural networks can also be applied in speckle-based optical cryptosystems to decrypt speckles for higher-level applications like face recognition.

It must be clarified that optical cryptosystems with high-security and fast-speed encryption have been investigated, and various applications in encrypting simple structural images (*e.g.*, characters, clothes, animals, *etc.*) have been demonstrated[12-15,25,26]. However, speckle-based optical cryptosystems for complex tasks, such as encrypted face recognition, have rarely been explored. The main challenge here is to decrypt images from rapidly changing optical speckles and to recognize faces from the decrypted images. Moreover, to achieve high accuracy in face recognition, decryption with high fidelity in key features and fine structures is required. In this work, we propose a scheme that utilizes optical speckles for face image encryption and a deep neural network for speckle decryption, and the decrypted images are then used for face recognition. The concept, as illustrated in Fig. 1, can be decomposed into three stages: first, face images are optically scrambled into speckles for encryption, which protects the data during transmission and storage; then, a neural network is trained to decrypt the face images with high fidelity from the ciphertext (*i.e.*, speckles); last, the decrypted images are compared with the known face encodings and recognized. In this cryptosystem, face images are encrypted into seemingly random speckles that are nearly impossible to be decrypted without the knowledge of the physical key (*i.e.*, the scattering medium) or the learned digital key (*i.e.*, the trained neural network). Moreover, only speckles but no face images are stored in the database to avoid any potential private information leakage. To the best of our knowledge, this is the first demonstration of a speckle-based optical cryptosystem for face recognition, and the accuracy in this study has reached more than 98%, which is applicable in a wide range of applications.

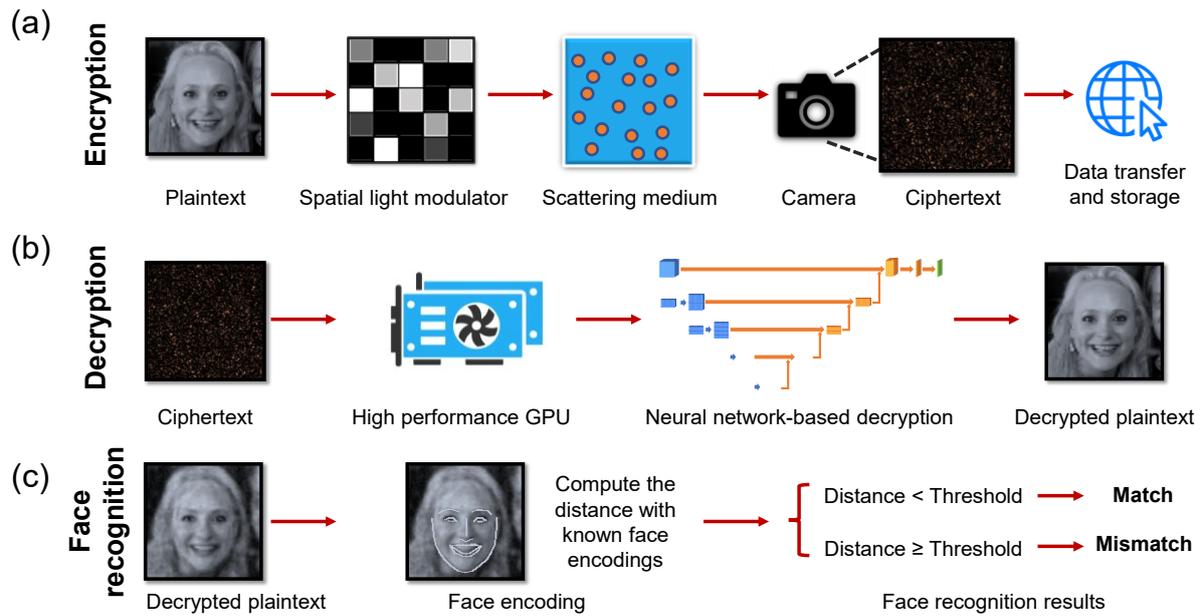

**Figure 1.** The flowchart of the proposed cryptosystem for face recognition. (a) Speckle encryption: face images (plaintext) are loaded on a spatial light modulator (SLM) to generate the corresponding speckles (ciphertext) when coherent light reflected by the SLM transmits through a scattering medium, which serves as the unique physical secret key. The ciphertext is safely transferred and stored via the cloud. No face images need to be kept in the database after encryption. (b) Learning-based decryption: a neural network is trained in advance to link the plaintext with the ciphertext. After training, new random speckle patterns (ciphertext) are directly fed into the neural network for decryption, and the decrypted face images are then utilized for face recognition. (c) Face recognition: the camera-recorded face images are encoded to unique 128-dimensional vectors of each known face image. After decryption, the face encoding distances between the decrypted images and the known face encodings are computed: if the encoding distance is less than a pre-set threshold, the face recognition result is "Match" (the same person), otherwise it is "Mismatch" (different people).

# Results

**Speckle-based encryption**

Fig. 2 shows the experimental optical setup for information encryption (see Methods for details). Face images from the "Flickr Faces High Quality" (FFHQ) database[27] are displayed on a phase-modulating spatial light modulator (SLM) to modulate the incident coherent light. Thus, the information of the face images (*i.e.*, plaintext) is carried by the wavefront modulated laser beam. Then, the modulated wavefront passes through a scattering medium (a 220-grit diffuser is used in this study) and is multiply scattered to form random speckles (*i.e.*, ciphertext), which are captured by a digital camera (FL3-U3-32S2M-CS, PointGrey, Canada). During encryption, which is the process of generating speckles, a MATLAB program synchronizes all devices to ensure each captured speckle pattern (*i.e.*, ciphertext) is paired with one exclusive face image (*i.e.*, plaintext) displayed on the SLM, as illustrated in Fig. 2. As seen, the ciphertext appears random and exhibits no direct relationship with the plaintext, and the mean Pearson correlation coefficient (PCC) between them is as low as 0.02.

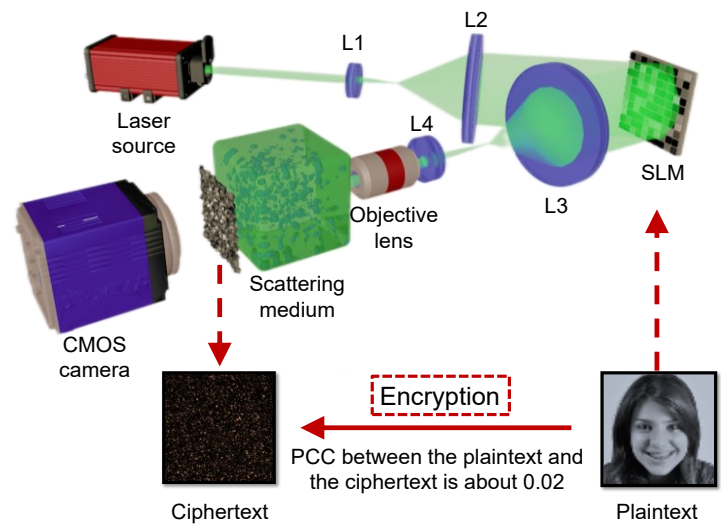

**Figure 2.** The optical setup for encryption. Face images (plaintext) are displayed on the SLM, which is illuminated by an expanded continuous coherent laser beam ($\lambda$=532 nm), generating speckles (ciphertext) through a scattering medium. The speckles are recorded by a CMOS camera, which is synchronized by a Matlab program to ensure one-to-one mapping with the displayed face image for network training.

**Learning-based decryption**

For information decryption from speckles, a neural network is constructed first. The structure of the neural network is shown in Fig. 3a, which is a U-Net[28] concatenated with a complex fully connected layer[20] and normalization layer, and the dimension of the filters in each layer is denoted in a format of length × height × amount (see Methods for details). Then the neural network is trained with 19,800 pairs of face images and their corresponding speckles (see Methods for details). The loss function used for training the neural network is

$$loss\ function = MSE(\hat{y}, y) - PCC(\hat{y}, y), \tag{1}$$

where $y$ is the ground truth and $\hat{y}$ is the predicted output from the neural network. Here, we adopt PCC to measure the overall similarity and mean square error (MSE) to measure the pixel-wise error. The experimental results of the neural network are shown in Figs. 3b-d. During the network training and evaluation, PCC gradually increases (Fig. 3d) and MSE gradually decreases (Fig. s1a), indicating increasing similarity between the decrypted images and the original plaintext. Especially, PCC becomes greater than 0.9 after 30 training epochs, suggesting high fidelity in decryption. In addition, we also measure other commonly used criteria, including the structural similarity index measure (SSIM) and the peak signal to noise ratio (PSNR), defined as Eqs. 2-5 in Methods. In Fig. 3b, four groups of exampled plaintexts, ciphertexts, and decrypted images during network testing are shown. The PCC, MSE, SSIM, and PSNR between the decrypted images and the original plaintexts are marked under the decrypted images. Overall, the average PCC, MSE, SSIM, and PSNR among all testing data (not included in network training) are 0.9422, 0.0083, 0.6884, and 21.25, respectively, demonstrating high accuracy of information decryption, which is essential for face recognition in the next stage. After network training, the plaintexts can be deleted from the cryptosystem to avoid privacy data leakage.

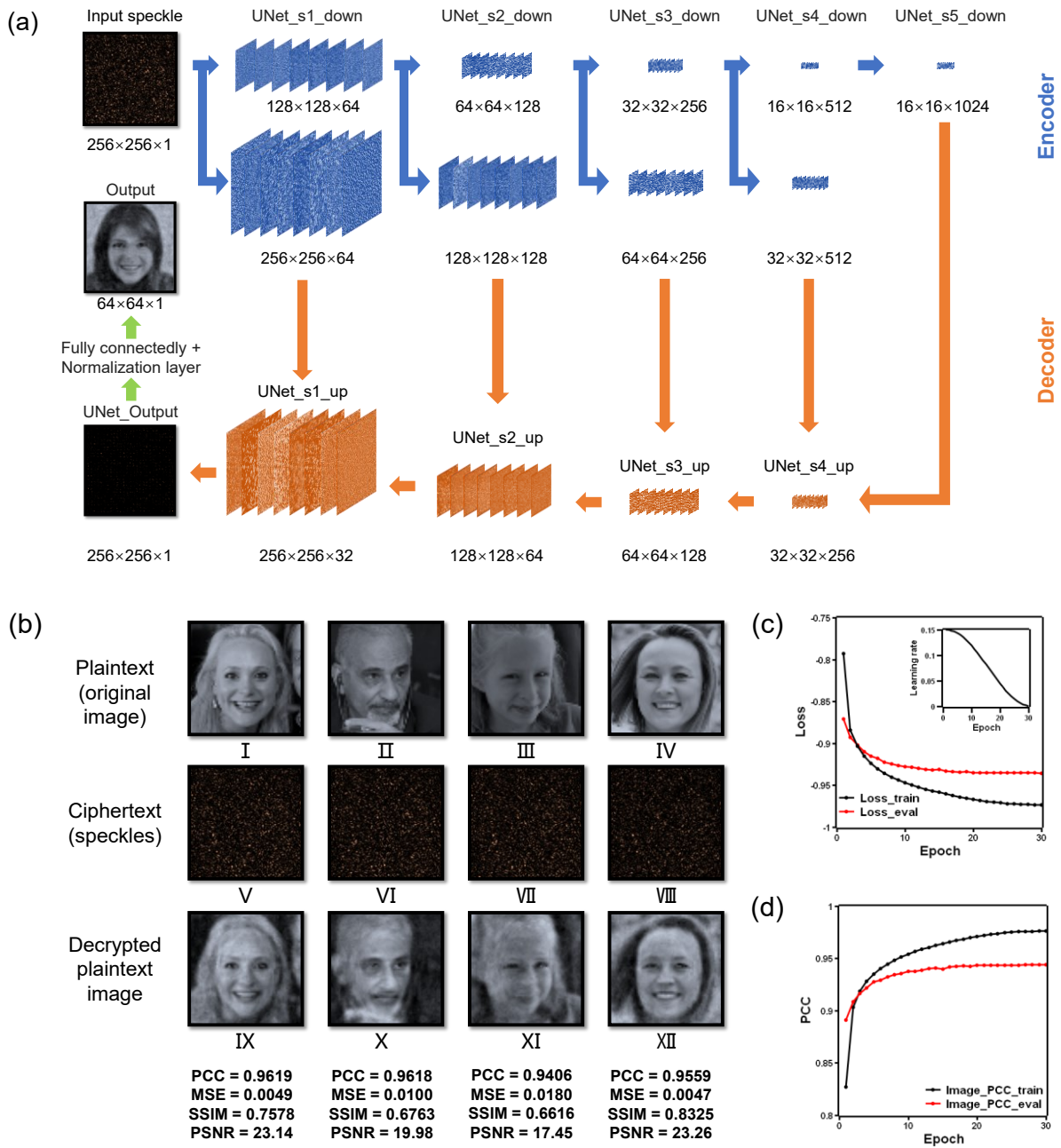

**Figure 3.** Neural network structure and the decryption performance. (a) Architectures of the neural network based on U-Net with an additional layer of a complex fully connected layer and normalization layer. The U-Net mainly contains 4 layers, with 4 down-sampling blocks for encoders (marked in blue) and 4 up-sampling blocks for decoders (marked in orange)[28]. The final outputs are face images decrypted from speckles, which are then used for face recognition. The dimensions of the filters are described as length × height × amount, and the filters shown here are visualized by inputting one speckle pattern into the neural network. (b) Four groups of exampled plaintexts, ciphertexts, and decrypted plaintext images during network testing. The ciphertexts are all from the same scattering medium, and the decrypted plaintext images are the results of inputting ciphertexts to the pre-trained neural network for decryption. The PCC, MSE,

SSIM, and PSNR between the decrypted and original images are marked under the corresponding decrypted images. (c) Loss function during training and evaluation. The inset shows the learning rate during network training. (d) The average PCC between the decrypted and original plaintexts during network training and evaluation.

Besides, the noise-resisting ability of the network is examined since noise always exists in experiment due to environmental disturbances, vibration, airflow, *et al*[21]. In our study, some computer-generated Gaussian noise with different standard deviations (*i.e.*, different noise amplitudes)[29] is added to the speckles for testing, and the decryption performance is updated with the pre-trained neural network. The results are given in Table s1 and Fig. 4a. In Table s1, the PCCs are all greater than 0.9 when the standard deviations of the noise are ≤ 0.3, which is consistent with what can be seen in Fig. 4a. The quality of the decrypted images deteriorates considerably when the standard deviation of the noise is ≥ 0.5 (*i.e.*, noise amplitude is half of the mean of the signal amplitude), and the face outline becomes indistinct. These results suggest that the neural network trained in this study can handle low and moderate noise conditions to the testing data, which is meaningful to the applicability of the method.

Furthermore, due to multiple light scattering and the conceptualized **infinite information channels**[16] within the scattering medium, it is hypothesized that the information of the plaintext is scrambled and distributes to the whole field of view (FOV) of the speckle pattern. Spatially, this speckle pattern could be large in practice, especially if the incident light is focused onto the front sample surface or the detection plane is far away from the sample. It is thus possible that only part of the speckle pattern is captured by the detection camera in experiments[30]. To study whether this factor may affect the performance, an additional group of experiments is conducted by using a quarter FOV of the speckle patterns for both network training and evaluation. That is, the dimension of the speckle patterns is reduced from 256×256 to 128×128 under the same spatial sampling condition. The experimental results are shown in Fig. 4b, Table s2, and Table s3. As seen, partial FOV leads to decryption results (Fig. 4b) that are very comparable to those obtained with larger FOV (Fig. 3b), confirming the hypothesis above. Such a non-point-to-point information mapping between the plaintext and the ciphertext is distinctive to most existing cryptosystems. It allows smaller speckle FOVs to be adopted in network training, evaluation, and testing, which can relieve the burdens of data collection, storage, and processing without compromising the decryption accuracy.

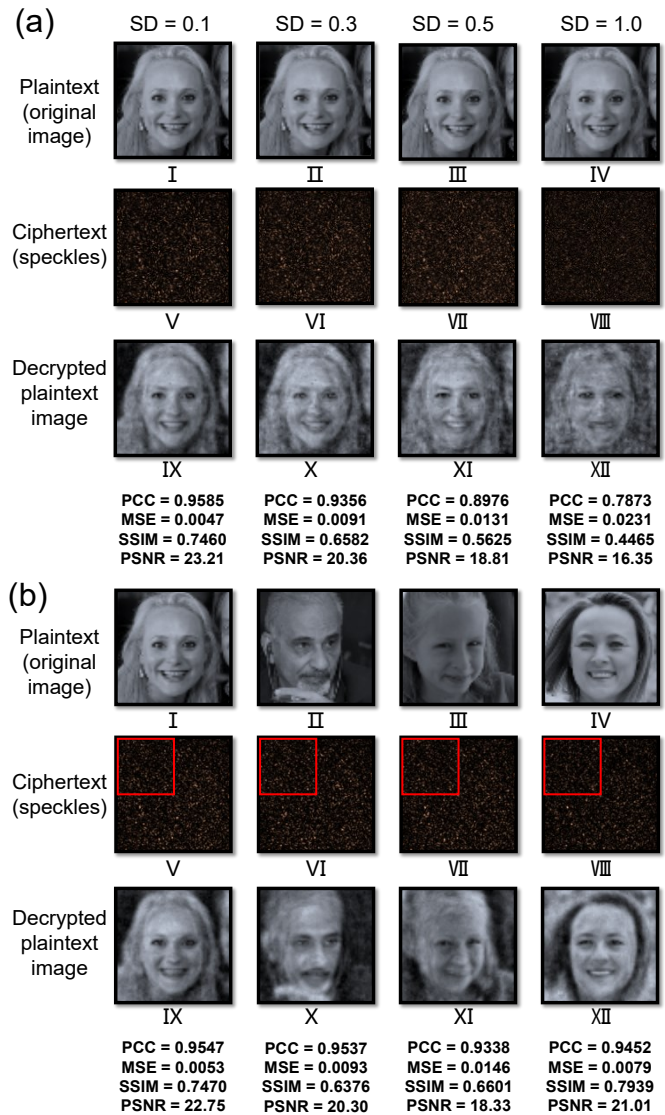

**Figure 4.** (a) Decryption performance with noisy speckles: the speckles with computer-generated random noise are fed into the pre-trained neural network for decryption. The noisy speckles and the corresponding decrypted images are marked with the corresponding noise standard deviation (SD) and similarity criteria. (b) Decryption performance with partial speckle patterns: only the top left corners (*i.e.*, quarter FOV, marked in red box) of the speckle patterns are used to train, evaluate, and test the neural network.

**Face recognition**

During decryption, we utilize PCC and other criteria to evaluate similarities. However, these criteria are not suitable for face recognition as they may be affected by many factors other than face features, such as image background, orientation, and expression of faces[31]. Therefore, at this stage the original and decrypted face images are further processed with an open-source

Python face-recognition library[32]. The neural network used for face recognition is based on ResNet[33], which is well-trained based on 3 million faces, with 99.38% accuracy on the Labeled Faces in the Wild benchmark [34,35]. The face recognition network encodes each face image with a unique 128- dimensional vector, which extracts the specific features of human faces, including eyebrows, eyes, noses, mouths, and cheeks. If the Euclidean distance[36] between two face vectors is lower than a pre-set threshold, two corresponding faces are defined as "Match" with each other; otherwise, they are defined as "Mismatch", as exampled in Fig. 5c. The commonly used pre-set threshold is 0.6 (for general situations) or 0.5 (for higher security scenes).

In our study, various thresholds between 0.5 and 0.6 are tested with decrypted face images illustrated in Fig. 3. As an example, the results of face recognition with a threshold distance of 0.6 are shown in Fig. 5. The key features of the original and decrypted face images from Fig. 3b are extracted by the face recognition neural network[32] and marked in the second row of Fig. 5a and b, respectively. As seen, most of these decrypted images appear akin to their corresponding original plaintext images (*e.g.*, image pairs I-V, II-VI, and III-VII, whose PCC are all more than 0.94) and hence are recognized as "Match". Note that, however, some image pairs seem visually alike, such as IV-VIII whose PCC ≈ 0.96, but are still recognized as "Mismatch" since the Euclidean distance is 0.61, being above the threshold of 0.6. Nevertheless, it shows that the face recognition library can extract key features and scale the differences between the decrypted and original face images.

Furthermore, we test the accuracy of face recognition. The 128-dimension face encodings from the decrypted images are compared with the corresponding encodings from the original face images, as shown in Fig. 5d. The results with different distance thresholds are shown in Table 1 and compared with other face recognition algorithms[37-41]. It is not surprising that different thresholds result in different recalls, precisions, and accuracies (Eqs. 9-12 in Methods). It can be observed that our accuracy reaches greater than 98% when the threshold is below 0.58. Compared with FaceNet and VGGFace, the method proposed in this work has higher accuracy and is therefore more suitable for practical applications[37-39]. Moreover, the precision is 100% when the threshold is set at 0.5, indicating high confidence during face recognition. However, the recall and F1 score obtained in this study are not as good as those from FaceNet and VGGFace and FaceNet, which can be attributed to the fact that there are more negative samples than positive samples in the data we use. The performance can be further improved by adjusting the threshold in face recognition according to the sample distribution in the dataset, or tuning the structure or parameters of the neural network.

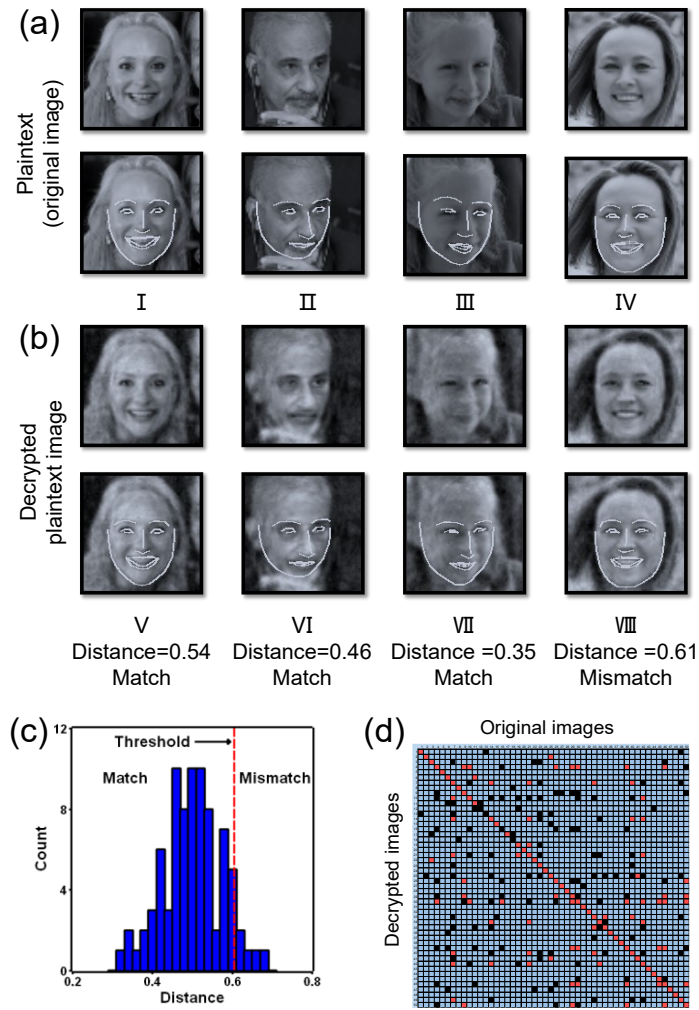

**Figure 5.** Face recognition results based on face images from FFHQ and the corresponding decrypted images from speckles. (a) The original face images (*i.e.*, plaintext) and their key features for face recognition. (b) The decrypted face images by feeding speckle patterns into the trained neural network and their key features. The face encoding distances between the decrypted and original face images with threshold = 0.6 are marked under the decrypted images. (c) Face encoding distances between the decrypted and original images in the test dataset. If the distance is less than or equal to the threshold = 0.6, the recognition result is "Match"; otherwise, it is "Mismatch". (d) The face recognition results of the decrypted images. True positives are marked in red, true negatives are marked in blue, while false positives and false negatives are marked in black.

Table 1. Face recognition results by our method and other algorithms with optimal thresholds.

|  | Threshold | Recall | Precision | Accuracy | F1 score |
|---|---|---|---|---|---|
| This work | 0.60 | 66.18% | 64.02% | 97.87% | 65.08% |
|  | 0.58 | 62.73% | 69.66% | 98.49% | 66.01% |
|  | 0.56 | 61.65% | 78.10% | 98.93% | 68.91% |
|  | 0.54 | 61.34% | 87.95% | 99.19% | 72.28% |
|  | 0.52 | 56.07% | 92.31% | 99.25% | 69.77% |
|  | 0.50 | 46.53% | 100.00% | 99.22% | 63.51% |
| FaceNet[38] | 0.90 | 96.42% | 100.00% | 98.21% | 98.18% |
| VGGFace[39] | 0.79 | 80.71% | 97.41% | 89.28% | 88.28% |
| OpenFace[40] | 0.47 | 16.42% | 95.83% | 57.85% | 28.04% |
| DeepFace[41] | 0.51 | 9.28% | 100.00% | 54.64% | 16.99% |

# Discussion and conclusion

In this study, a speckle-based optical cryptosystem is proposed, implemented, and demonstrated, by exploiting a ground glass scattering medium as the physical secret key to generate speckle patterns that uniquely encrypt information. As for a cryptosystem, security is the topmost concern, we will discuss the security of the proposed method from three aspects.

**Length of the secret key**

The equivalent key length of the scattering medium can be modelled by the transmission matrix, whose dimension in this work is (256×256) × (64×64), and each element is 64 bits (for complex float numbers) in computer. Thus, the digital key of this cryptosystem is of length 64 × [(256×256) × (64×64)] = $1.72×10^{10}$ bits (*i.e.*, 17.2 gigabit), which is enormous for brute force attacks even with a quantum computer. In comparison, for purely software-based encryption approaches, such as Advanced Encryption Standard (AES)[5] and Compression Friendly Encryption Scheme (CFES)[42], the digital cryptosystems are all based on matrix manipulations. As the size of the matrix (*i.e.*, digital secret key length) increases, more multiplicative manipulations are needed, and the computational complexity grows exponentially. Therefore, to balance the computational efficiency and security, the digital secret key lengths in digital cryptosystems are usually limited to hundreds of bits. However, in our speckle-based encryption process, no mathematical algorithms are involved, so the computational burden can be ruled out during encryption and users can achieve high security without compromising encryption speed. Note that, when it comes to decryption, both optical and software-based cryptosystems involve a large amount of computation. Fortunately, these decryption processes can be accelerated by using a high-performance GPU.

**Unclonable physical secret key**

As for the optical setup, it is nearly impossible to generate the same speckles with a different scattering medium (*i.e.*, the physical secret key), in which the scatterers are randomly distributed, and the propagation behavior of photons is very complicated. Therefore, compared with existing digital encryption matrix-based approaches (*i.e.*, relays only on digital secret keys)[43], it is nearly impossible to duplicate the scattering medium to crack the cryptosystem[13], except for a self-defined medium such as a metasurface[44-45]. Therefore, the speckles can be viewed as nearly unclonable, and the decryption process is exclusive to the quantification of the scattering medium, *i.e.*, a DNN trained with ciphertext (*i.e.*, speckles) as the input and plaintext as the output. If speckles generated from another scattering medium (*i.e.*, wrong physical secret keys) are input to the pre-trained neural network for decryption, as shown in Fig. 6, the decrypted results (XIII to XVIII) are obscure and very different from the plaintext (I to VI). As a result, the decrypted images cannot be used for face recognition and thus the security of the proposed system can be guaranteed.

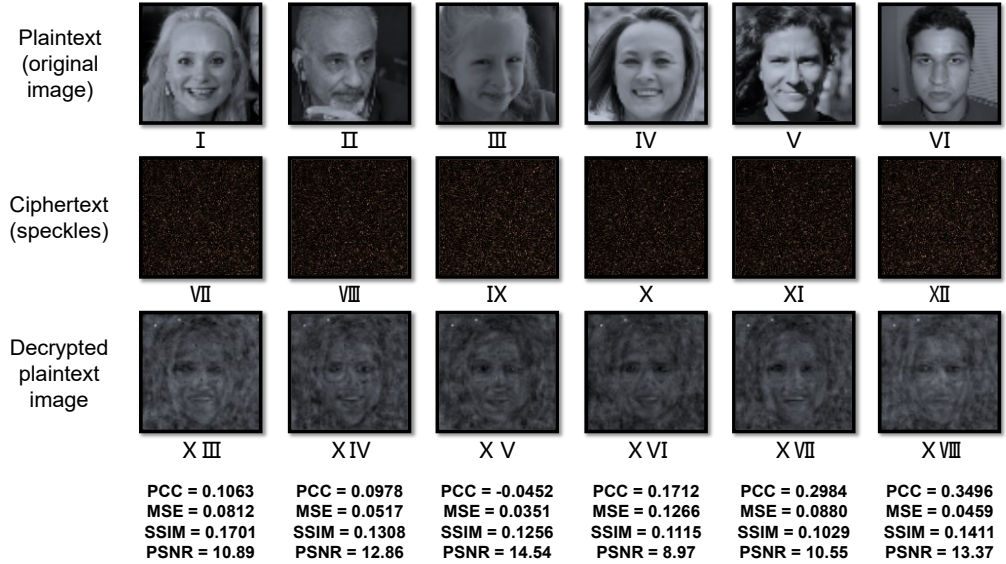

**Figure 6.** Wrong physical secret key attack: same plaintext images are used, but another scattering medium is utilized to generate the speckles (*i.e.,* ciphertext), which are input to the pre-trained neural network to yield the decrypted plaintext images. The PCC, MSE, SSIM, and PSNR between the decrypted and the corresponding original face images are marked.

**Uniqueness of the optical setup**

Under extreme situations when hackers have obtained the scattering medium (*i.e.,* the physical secret key), to produce the same speckle patterns, the error in duplicating the optical system alignment and the light-medium interaction should be within the optical wavelength scale[46]. That is, the optical setup ensures that the interaction between the light and medium is hard to be reproduced due to the 'narrow' range (~milliradians for tilt and submicron for shift) of the 'memory effect'[46]. Beyond the memory effect, it is theoretically impossible to reconstruct images from speckles or speckle autocorrelations[47].

**Other advantages**

The intervention of optics further boosts the efficiency of encryption (*i.e.,* at the speed of light) and overwhelms the software-based cryptosystems. Therefore, optical solutions, including the proposed speckle-based method and Double Random Phase Encryption (DRPE) method, can enable highly efficient encryption and generate high-dimensional secret keys[8]. Notably, compared with DRPE, the proposed method is advantageous due to its simpler optical design. DRPE requires two SLMs in the optical setup since the information is encrypted by two random phase masks[13]. In our cryptosystem, the encryption can be performed with a scattering medium only. This not only facilitates the integration with other systems, but also reduces the cost of the cryptosystem. The most expensive component in the current system is the SLM, which is only responsible for loading the images and is indeed replaceable in practice since direct illumination of human faces can be used as input images for the cryptosystem. As a result, the cost of the proposed cryptosystems becomes comparable to the software-based cryptosystems, which only require computers for encryption and decryption.

When it comes to system latency, although well-known edge computing can help to recognize face images and protect privacy through computing in cloudlets, its scalability is refrained by the computing power, leading to applications of limited database[48]. In comparison, the proposed light-based system can achieve fast encryption speed and high scalability. Moreover, with the development of high throughput communication networks, such as 5G, the latency of the proposed system is also comparable to edge computing-based face recognition[49].

**Conclusion**

With the proposed speckle-based optical cryptosystem, the encrypted private data (*e.g.*, face images) is difficult to crack and reduces the risk of information leakage. The speckle-based optical cryptosystem is suitable for practical applications due to its high security, fast speed, low cost, insensitivity to the field of view, as well as immunity to low and moderate noise to the ciphertexts. That said, the accuracy of face recognition can still be further improved by constructing more complex neural networks that lead to an all-speckle-based optical cryptosystem for decryption and face recognition[50,51], where there is no need to decrypt optical speckles to face images. Moreover, to further enhance the security of the encryption processes, multi-channel laser diffraction by high-dimensional scattering media can be adopted to increase the speckle randomness. On the other hand, binary speckles can be used to reduce data storage space and increase data transmission speed[52]. Collectively, although this study contains only proof-of-principle demonstration for face encryption and recognition, we believe that with further optimization the proposed speckle-based optical cryptosystem may find or inspire wide applications in high-security information encryption and decryption.

## Methods

### Optical setup

The experimental setup during speckle encryption is shown in Figure 2. First, the human face images are loaded onto the SLM (HOLOEYE PLUTO VIS056 1080p, German). The human face images used here are taken from the thumbnails of the Flickr-Faces-HQ (FFHQ) database, a dataset of human face images. The original FFHQ database contains 70,000 images, from which we select the first 20,000 images for demonstration in our study. The beam of a continuous wave 532 nm laser (EXLSR-532-300-CDRH, Spectra-Physics, USA) is expanded by a 4-f system (L1 and L2 in Fig. 2) so that the SLM is fully illuminated to modulate the incident laser. In experiments, the resolution of the SLM is 1920×1080, and the 128×128 thumbnails are up-sampled to 1024×1024 and loaded onto the SLM to fully utilize the modulation capability. The intensity of the image on the SLM is then converted to a phase delay (0 to 2π). Finally, the wavefront-modulated beam light is focused by an objective lens (RMS20X, Olympus, Japan) onto and propagates through a scattering medium (220-grid, DG10-220-MD, Thorlabs, USA). In experiments, 20,000 images are sequentially loaded onto the SLM, and the corresponding speckle patterns are captured by a CMOS camera (FL3-U3-32S2M-CS, PointGrey, Canada) with a resolution of 256×256.

### Training dataset

The speckles used as the network input are 256×256 speckle images captured by the CMOS camera, and the images from FFHQ used as the network output are 64×64 images down-sampled from the FFHQ dataset (128×128) to avoid using up the GPU memory. These resolutions are chosen to make full use of the experimental setup and achieve high fidelity image decryption. The size of the training dataset is 19,800, leaving 100 datasets for evaluation and 100 datasets for testing, respectively. Before the speckle data is input to the neural network, the input data are linearly normalized from 0 to 1 for better neural network performance[53].

### Neural network for decryption

The detailed structure of the neural network for decryption is shown in Fig. 3a. Overall, the architecture of the neural network is based on commonly used U-Net[28] with an additional complex fully connected layer[20] and a normalization layer. The encoders in the U-Net contain 4 down-sampling blocks and the decoders in the U-Net contain 4 up-sampling blocks. In addition, the fully connected layer is based on complex numbers. In Fig. 3a, the blue arrows and filters represent the encoders in the U-Net, and the orange arrows and filters represent the decoders in the U-Net. The encoder tends to extract low-dimensional features from the speckles and encodes them. The decoder then tends to extract high-dimensional features and decodes them[28]. As a result, the encoder and decoder-shaped neural network can extract features of different dimensions. The fully connected layer is used as the last layer to transform extracted features into images. The normalization layer limits the output range to [0,1]. At last, the final output is the face images decrypted from random speckles, which are then used for face recognition.

During neural network training, the optimizer used in training the neural network is stochastic gradient descent (SGD)[54], and the learning rate is 0.15, with cosine annealing. During the experiments, we train the neural network for 30 epochs, and the neural network is then tested. The software framework used is Pytorch 1.8.0 with Python 3.7.6 and CUDA 10.1 for GPU acceleration. The hardware we used is Dell Precision Tower 5810 with Intel Xeon E5-1650 V3 CPU, 64 GB RAM, and Nvidia GeForce RTX 2080Ti 11GB GPU. During the training, one epoch takes about 30 minutes, and the whole training process takes about 15 hours.

**Image similarity criteria**

During neural network training and testing, we use PCC, MSE, PSNR, and SSIM as the image similarity criteria, which are defined in Eqs. 2-5:

$$PCC = \frac{mean[(y-mean\,(y))\times(\hat{y}-mean\,(\hat{y}))]}{std\,(y)\times std\,(\hat{y})} \tag{2}$$

$$MSE = mean\,[(\hat{y}-y)^2] \tag{3}$$

$$PSNR = 20 \times \log_{10}\frac{max\,(\hat{y},y)}{\sqrt{MSE}} \tag{4}$$

$$SSIM = l(\hat{y},y) \times c(\hat{y},y) \times s(\hat{y},y) \tag{5}$$

$$l(\hat{y},y) = \frac{2\times mean(y)\times mean(\hat{y})+c_1}{mean(y)^2+mean(\hat{y})^2+c_1} \tag{6}$$

$$c(\hat{y},y) = \frac{2\times std(y)\times std(\hat{y})+c_2}{std(y)^2+std(\hat{y})^2+c_2} \tag{7}$$

$$s(\hat{y},y) = \frac{cov(y,\hat{y})+c_3}{std(y)\times std(\hat{y})+c_3} \tag{8}$$

In the equations above, $y$ and $\hat{y}$ are the original and decrypted images, respectively; $mean(y)$ and $mean\,(\hat{y})$ are the mean values of $y$ and $\hat{y}$, respectively; $std(y)$ and

$std\ (\hat{y})$ are the standard deviation of $y$ and $\hat{y}$, respectively; $cov(y, \hat{y})$ is the covariance of $y$ and $\hat{y}$; $c_1$, $c_2$, $c_3$ are three very small constants ($10^{-5}$) to prevent division by 0 in SSIM[55]; $l(\hat{y}, y)$ is the luminance similarity; $c(\hat{y}, y)$ is the contrast similarity; $s(\hat{y}, y)$ is the structure similarity. Among these criteria, we only use MSE and PCC in the loss function during network training, and other criteria are just used during network evaluation and testing.

**Face recognition criteria**

The decrypted images are input to an open-source face recognition program for face recognition[32]. Before testing the neural network, some images with sunglasses and babies are excluded since some of their facial key points are ambiguous. The most important criterion during network testing is face recognition accuracy. First, the face recognition program encodes each face image with one special 128-dimension encoding[32], which takes less than 1 second. Then, our target is that if the Euclidean distance[36] between the encoding vectors of two original images are smaller than the preset threshold (indicating that they are the same person), the distances between the two corresponding decrypted images are also expected to be smaller than the preset threshold, indicating that the person in the decrypted images and the original images are "match". Here, mainstream computers to date (e.g., Xeon E5-1650 V3 with 6 cores in experiments) can handle more than 10,000 face encoding distances within 1 second.

The encodings of the decrypted images are also compared with each encoding of the original images. If the two original images' encoding distances are smaller than the preset threshold, the two samples are treated as positive samples. And if the corresponding two decrypted images' encoding distances are also smaller than the preset threshold, the results are true positives, otherwise they are false negatives. On the contrary, if the two original images' encoding distances are larger than the preset threshold, the two samples are treated as negative samples. And if the corresponding two decrypted images' encoding distances are also larger than the preset threshold, the results are true negatives; otherwise, they are false positives. During network testing, precision, recall, F1-score, and accuracy are used to evaluate the performance, as defined in Eqs. 9-12.

$$Recall = \frac{True\ Positive}{True\ Positive + False\ Negative} \qquad (9)$$

$$Precision = \frac{True\ Positive}{True\ Positive + False\ Positive} \qquad (10)$$

$$Accuracy = \frac{True\ Positive + True\ Negative}{True\ Negative + True\ Positive + False\ Negative + False\ Positive} \qquad (11)$$

$$F1\ score = 2 \times \frac{Precision \times Recall}{Precision + Recall} \qquad (12)$$

As one person might be recognized as two different people while two different people should not be recognized as the same person, accuracy is more meaningful than the other three criteria in this study.

# Acknowledgements

This work was supported by National Natural Science Foundation of China (NSFC) (81930048, 81627805, 81671726), Guangdong Science and Technology Commission (2019A1515011374, 2019BT02X105), Hong Kong Research Grant Council (15217721, 25204416, R5029-19), Hong Kong Innovation and Technology Commission (GHP/043/19SZ, GHP/044/19GD, ITS/022/18), and Shenzhen Science and Technology Innovation Commission (JCYJ20170818104421564). The authors would like to thank the Photonics Research Institute and University Research Facility in Big Data Analytics of the Hong Kong Polytechnic University for facility and technical support.

# Conflict of Interest

The authors declare no conflict of interest.

# Author Contributions

Qi Zhao, Huanhao Li, and Zhipeng Yu contributed equally to this work. Qi Zhao, Huanhao Li, Zhipeng Yu, Jie Tian, and Puxiang Lai conceived the idea. Qi Zhao, Huanhao Li, and Zhipeng Yu conducted the experiments and processed the data. Qi Zhao, Huanhao Li, Zhipeng Yu, and Chi Man Woo wrote the manuscript. Jie Tian and Puxiang Lai supervised the project. All members contributed to the discussion of the results and proofreading of the manuscript.

## Data Availability Statement

The data that support the findings of this study are available from the corresponding author upon reasonable request.